\documentclass[a4paper,12pt,reqno]{amsart}
\usepackage[T1]{fontenc}
\usepackage[utf8]{inputenc}
\usepackage[english]{babel}
\usepackage[left=25mm,right=25mm,top=35mm,bottom=35mm]{geometry}
\usepackage{amsmath,amssymb,amsthm}
\usepackage[scr=rsfs]{mathalpha}
\usepackage{pgfplots}
\usepackage{pgfplotstable}
\usepackage[margin=0pt,font={small}]{caption}
\usepackage{booktabs}
\usepackage{color,colortbl}
\usepackage[colorlinks=true,allcolors=blue,breaklinks]{hyperref}

\newcommand{\R}{\mathbb{R}}

\newcommand{\LL}{\mathcal{L}}

\newcommand{\RR}{\mathcal{R}}
\renewcommand{\SS}{\mathcal{S}}

\def\vfld#1{\vec{#1}}
\def\ifiss{{\sc ifiss}}
\def\Ra{{Ra}}
\def\Re{\RR}

\definecolor{myBrown}{rgb}{0.6 0.4 0.2}
\definecolor{myOrange}{rgb}{1.0 0.6 0.2}
\definecolor{myLightGray}{RGB}{235,235,235}
\definecolor{myViolet}{RGB}{153,50,204}

\makeatletter
\def\@seccntformat#1{%
  \protect\textup{\protect\@secnumfont
    \ifnum\pdfstrcmp{subsection}{#1}=0 \bfseries\fi
    \csname the#1\endcsname
    \protect\@secnumpunct
  }%
}
\makeatother
\newcommand*\patchAmsMathEnvironmentForLineno[1]{%
  \expandafter\let\csname old#1\expandafter\endcsname\csname #1\endcsname
  \expandafter\let\csname oldend#1\expandafter\endcsname\csname end#1\endcsname
  \renewenvironment{#1}%
     {\linenomath\csname old#1\endcsname}%
     {\csname oldend#1\endcsname\endlinenomath}}%
\newcommand*\patchBothAmsMathEnvironmentsForLineno[1]{%
  \patchAmsMathEnvironmentForLineno{#1}%
  \patchAmsMathEnvironmentForLineno{#1*}}%
\AtBeginDocument{%
\patchBothAmsMathEnvironmentsForLineno{equation}%
\patchBothAmsMathEnvironmentsForLineno{align}%
\patchBothAmsMathEnvironmentsForLineno{flalign}%
\patchBothAmsMathEnvironmentsForLineno{alignat}%
\patchBothAmsMathEnvironmentsForLineno{gather}%
\patchBothAmsMathEnvironmentsForLineno{multline}%
}
\usepackage[mathlines]{lineno}

\usepackage{fancyhdr}
\advance\footskip0.5cm
\pgfplotsset{
tick label style={font=\tiny},
legend style={font=\tiny},
xlabel style={yshift=+0.5ex},
ylabel style={yshift=-1.0ex}
}
\definecolor{otherblue}{rgb}{0,0.3,0.6}
\def\rbl#1{\textcolor{black}{#1}}

\newcommand\rems[1]{{\color{otherblue}#1}}

\title{Machine learning for hydrodynamic stability }
\author{David J. Silvester}
\address{Department of Mathematics, University of Manchester, Oxford Road, Manchester M13 9PL, UK}
\email{d.silvester@manchester.ac.uk}
\thanks{{\em Acknowledgements.}
\rems{This work was supported by EPSRC grants {EP/W033801/1} and  {EP/Y028783/1}.
The authors would like to acknowledge the support provided  by Research IT and the use of the 
Computational Shared Facility at The University of Manchester.}
}
\date{\today}

\begin{document}

\begin{abstract}
A  machine-learning strategy for investigating the stability of fluid flow problems is proposed herein. 
The goal  is to provide a simple yet robust methodology to find a nonlinear mapping  from the parametric space to an
indicator representing the probability of observing a bifurcated solution. The computational procedure  is demonstrably robust and does not require parameter tuning. The essential feature  of  the strategy is that the computational solution of the Navier--Stokes equations is  a reliable proxy for laboratory experiments investigating sensitivity to flow parameters. The applicability of our
probabilistic bifurcation detection strategy is demonstrated by an investigation of  two classical examples of flow instability associated with thermal convection.  The codes used to generate and process the labelled data are available on GitHub. 
\end{abstract}


\maketitle
\thispagestyle{fancy}

\section{Introduction} \label{sec:intro}

The issue of hydrodynamic stability has been extensively studied over the last century. Excellent books on the topic include 
\cite{joseph76}, \cite{landau87},  \cite{drazin81} and  \cite{schmid01}. 
Linearizations for small perturbations of a base  (steady) flow are the starting point for developing mathematical theory 
and often provide results that match with laboratory  experiments. Classical linear stability analysis assumes exponential 
behaviour of infinitesimal perturbations and leads to characteristic  eigenvalue problems that can be analysed
theoretically (following in the  footsteps of Lord Rayleigh). They can also be explored computationally by  discretising
the governing partial differential equations (PDEs) linearised about the base flow and solving a high-dimensional 
eigenvalue problem to track the critical eigenvalues. See \cite{cliffe00} for a comprehensive review and
 \cite[Chap.\,6]{wubs23} for a  contemporary treatment. A practical  issue  is that classical eigenvalue analysis provides 
no information about the size of  asymmetric perturbation needed \rbl{to  initiate}  a  transition to an asymmetric
flow profile from an unstable symmetric flow pattern. In this work,  the  influence of the {\it perturbation magnitude} will be
assessed.  The motivation for this  is to  provide insight into the relative difficulty of  assessing flow stability by
laboratory experiment.
The concept of  ``learning from data''  has become mainstream over the past decade with activity in all areas of science. 
Quoting from \cite{strang19}, the central goal of machine learning is to construct a function that classifies the training 
data correctly,  so it can generalise to unseen test data. Assessing the effectiveness of a  rudimentary neural
 network  in correctly classifying hydrodynamic stability will be the focus of this study.
 
 There is a bugeoning literature on machine learning techniques for bifurcation problems. State-of-the-art machine learning 
 techniques, such as those discussed by \cite{szep21} or \cite{mogharabin25}  can efficiently determine multiple bifurcations
 in cases where the underlying dynamical system is  low-dimensional.  Julia software tools such as ModelingToolkit.jl can 
 autodifferentiate ODE systems and  when interfaced with BifurcationKit.jl, can map from right-hand sides to continuation and 
 diagram construction. However, as noted in   \cite{szep21}, these sohisticated detection strategies may not scale effectively 
 for partial differential  equations  where a large number of states  arises from discretisation of the spatial variables. 
 Other specialised machine learning  methods  such as those discussed in  \cite{celletti25} can be used  to classify different 
 types of motions (regular or chaotic)  from  a time series or even be used to approximate Lyapunov exponents using neural 
 networks \cite{mayora-cebollero25}. Both  of these tasks are  significantly harder  than the binary classification distinguishing
between  bifurcated and non-bifurcated states that  is the focus of this study.

 Starting  from the pioneering paper \cite{raissi20}, problems from fluid dynamics have featured prominently in 
papers exploring  machine learning of solutions of partial differential equations. Efforts to train physics-informed neural networks 
(PINNs) in the context of fluid flow modelling have met with mixed success to date however. \cite{wang23} identify 
the Stokes and Navier--Stokes equations as challenging problems that highlight some of the most prominent difficulties in 
training PINNs. Recent work \cite{pichi23} builds on earlier work \cite{chen21} in constructing a 
machine-learning strategy for learning the steady-state bifurcation structure of  flow problems. 
The developed strategy  is limited by a computationally expensive offline stage whereby a reduced basis set is computed 
from a singular-value  decomposition of a (high-dimensional) snapshot matrix. 
The associated error in this dimensional reduction is difficult to quantify and limits 
 the achievable accuracy when modelling fluid flow (because  the decay of the singular values is so slow). In contrast, the 
 methodology developed herein is based on direct simulation of the flow problem, where the effect of discretisation 
 error can be  monitored to ensure that it does not influence the resulting flow classification. Neural network training is
 compared with  Gaussian process regression in a recent paper \cite{sousedik22} assessing the 
 linear stability of  two benchmark  flow problems. The issues associated with the computation and careful tracking of 
 eigenvalues are eschewed in our bifurcation detection strategy.
 
 
 Our stability classification strategy involves two components. First and foremost, we need an accurate discretisation 
 strategy for the governing PDEs modelling the time-dependent evolution of the flow. The second component is to specify a 
 bifurcation detection strategy and assign a tolerance for  labelling the training data. Precision is not so critical in this
 however---our experience is that the resulting neural network classification results are relatively insensitive to the choice of 
tolerance. A description  of the particular machine-learning strategy that we employ  can be found in section~2.
 
 Regarding the discretisation strategy, a standard method-of-lines approach that combines a stable finite 
 element spatial approximation together with a similarly accurate implicit time discretisation is adopted herein. One feature 
 distinguishing our strategy from that of others  is the use of self-adaptive time stepping that is designed to ``follow the 
 physics''.  The benefit of this is  that instability of the computed flow does not need to be explicitly excited. In singular cases 
 (multiple steady-state flows) the onset of instability is triggered by computer round-off  much as it would be triggered in 
 physical experiments by geometrical imperfections in the laboratory apparatus. The upshot however, is that stable  linear
 solver strategies are essential if we are to respect the singularity or ill-conditioning of the linearised  algebraic equation 
 systems. These issues are explored in the context of flow problem with a symmetry-breaking pitchfork flow bifurcation in 
 section~3.  The applicability of our bifurcation detection strategy is demonstrated in section~4, where we present results for 
 two examples of flow instability associated with thermal convection that have been partially studied previously.
 
 The use of a neural network surrogate to model
the stability of flow configurations generated by direct simulation of the  Navier--Stokes equations
 is completely novel as far as we are aware. The methodology will be shown to offer new insights into the 
 assessment of the stability of the specific  flow problems that feature in this study.
 
 \section{Bifurcation classification using a neural network} \label{sec:logistics}

The two distinct  flow problems considered in this work are defined on a bounded spatial domain $\rbl{\Omega \subset \R^2}$ and 
a temporal domain $\tau=(0,\infty)$.  Given a time-dependent PDE operator $\rbl{\mathscr{ L}}$ 
(typically representing the Navier--Stokes equations) and a vector  $\lambda\in \R^n$ of parameters, 
the goal is to explore the bifurcation structure of  flow solutions 
$u({\vec x},t;\lambda)$ \rbl{and $p({\vec x},t;\lambda)$} satisfying
\begin{align}
\label{eq:pde:abstract}
 \rbl{\mathscr{ L} (u({\vec x},t;\lambda), p({\vec x},t;\lambda))}
= 0
& \quad\text{in $\Omega\times \tau$}, 
\end{align}
\rbl{starting from a quiescent flow  at time $t=0$, combined with physically realistic boundary conditions on 
$\partial\Omega\times \tau$.} 

 The characterisation of the bifurcation structure of a given flow problem is a two-stage procedure. In the 
first stage, approximations  $u_h$ of solutions $u$ will be computed by discretising~\eqref{eq:pde:abstract} 
in space (using mixed finite element approximation) and in time (using adaptive time stepping driven 
by local error control), by sampling  parameter vectors $\lambda^j$ in the neighbourhood 
of the bifurcation boundary.\footnote{This step requires a priori knowledge of the bifurcation 
phenomenon. While this can often be guessed by physical considerations or  from laboratory 
experiments, determining  suitable sample points  would pose a challenge  in cases where 
the system behaviour is totally unknown.}
A classification test is then applied to the sampled data to determine whether the 
computed flow is representative of the base flow or else has undergone a bifurcation.  
The result of this is a set of labelled data
\begin{equation}
\SS = \{ (\lambda^j,\ell^j) \}_{j=1}^{{m}},
\end{equation}
where  the label $\ell^j$ is either 0 or 1.  A distinguishing feature of our
time stepping strategy is that a bifurcated  flow solution $u_h^j$  typically {\sl evolves} from the 
associated base state (this depends on the underlying bifurcation structure). Small perturbations of the 
flow solution are not explicitly introduced into the time stepping process. This supports our assertion 
that the computer can be viewed as a proxy for a corresponding laboratory experiment investigating 
sensitivity to flow parameters.

The second stage of our procedure is the generation of a surrogate
model of the bifurcation structure.  We will confirm that this can be done simply and effectively
by training a shallow neural network to classify the data.  The machine-learning 
ingredients will be  elementary: a shallow network will prove to be  perfectly adequate 
and we do not  need to carefully  tune any  hyperparameters.  A random division of  the  data in $\SS$
into a  training dataset and a test dataset will demonstrate that the machine-learned models of  
bifurcation structure are statistically valid.  A representative 
surrogate generated with  fixed PDE discretisation parameters can also be validated a posteriori, by 
sampling parameters close to the bifurcation boundary and comparing the model predictions with the labelled
data that is generated for the same parameter values using a refined finite element mesh or  a 
smaller time accuracy tolerance. 

The neural  network architecture was fixed throughout this study. 
The input to the network is the parameter vector $\lambda$ (two-dimensional in all cases
discussed later) and the final output is a prediction of the corresponding label $\ell$. 
The network has a single  hidden layer consisting  of 32 neurons.  Changing the width
to 16  or  to 64 neurons  does not affect the results.
The activation  function is the standard  sigmoid.  The  output from the 
hidden layer  is  a two-dimensional vector  $[z_1,z_2]$ that is mapped onto a vector of
 probabilities $p=[p_1,p_2]$ using  a numerically stable softmax function 
\begin{equation}\label{stablesoftmax}
z_* = \max\{z_1,z_2\},\qquad
p_j = {e^{z_j-z_*} \over e^{z_1-z_*} + e^{z_2-z_*}}, \quad j=1,2.
 \end{equation}
The resulting output vector  will be interpreted  as the probability of computing 
{a bifurcated} solution ($p_1$)  or {the base} solution ($p_2$).  The training of the neural
network is accomplished by comparing the output vector $p$ with the 
labelled output vector  
\begin{equation} \label{labelvector}
y=
\begin{cases} 
     [0,1] & \hbox{ if  } \ell^j=0, \\
     [1,0] & \hbox{ if  } \ell^j=1.
\end{cases}
\end{equation}
using a  cross-entropy loss function ({\sl logistic regression}) 
with the loss functional 
\begin{equation} \label{likelihood}
\LL(y,p) = - y_1 \ln(p_1) - y_2 \ln(p_2).
\end{equation}

 \begin{figure}[!thp]
\centering
\includegraphics[width = 0.9\textwidth]{{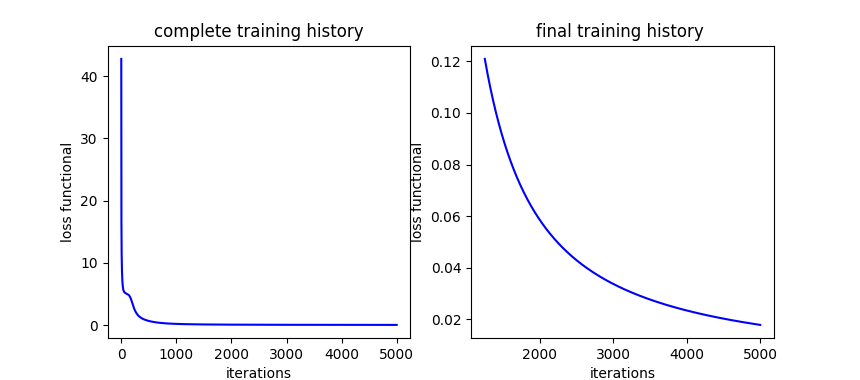}}
\caption{Network training history for 105 data points (symmetric channel flow bifurcation).}
\label{fig:train}
\end{figure}

In all cases discussed in later sections, the training of the network proved to be  robust with respect to
the initialisation of the weights and biases and did not require bespoke parameter
tuning.\footnote{The raw parameter data does need to be normalised however. This was done
by  subtracting the mean and dividing by the standard deviation of the individual components.} 
A typical training run with ${m}=105$ data points with a fixed learning rate $\alpha=0.15$ 
 is shown in Figure~\ref{fig:train}.  The optimisation algorithm consisted of  5000 passes 
 through the entire dataset,  taking a  gradient descent step after every data point with
 gradients computed analytically by back-propagation.
 Looking closely  at the convergence history,  we can see that the output of the loss functional
 decreased monotonically with a reduction of  an order of magnitude 
 in the first 20 steps followed by a monotonic reduction by a further factor of 2 (50\%)  after 200 steps. 
 The monotonic decrease of the loss  functional is increasingly  slow thereafter: the relative reduction achieved in the  
 final 4000 steps was an additional  factor of 10.  The $\ell_2$ norm of the gradient vector when the
 training run ended was $3.16\times 10^{-35}$.

\section{Validation of the  classification strategy} \label{sec:validation}

\begin{figure}
\centering
\begin{picture}(300,75) (24,80) 
\put(35,57){\includegraphics[width=.6\textwidth, height=.2\textwidth,clip]{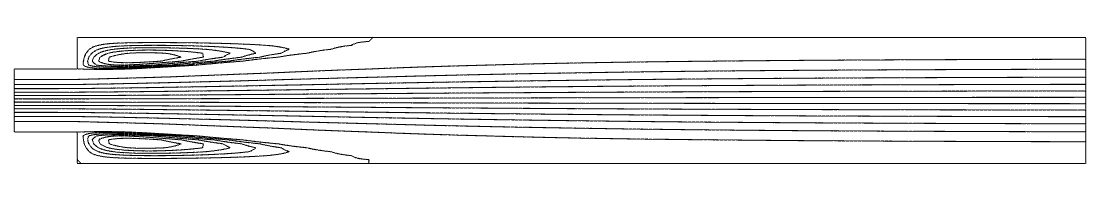}}
\put(32,105){\vector(0,1){12}}
\put(32,105){\vector(0,-1){12}}
\put(26,103){{\scriptsize 1}}
\put(44,145){\line(1,0){10}}
\put(44,143){\line(0,1){4}}
\put(54,143){\line(0,1){4}}
\put(48,147){{\scriptsize 1}}
\put(315,105){\vector(0,1){24}}
\put(315,105){\vector(0,-1){24}}
\put(318,103){{\scriptsize $2$}}
\put(183,145){\vector(1,0){119}}
\put(172,145){\vector(-1,0){119}}
\put(173,143){{\scriptsize $16$}}
\end{picture}
\caption{Symmetric step domain and computed  flow solution for $\Re=212.8$.} 
\label{step-figure-I}
\end{figure}

The effectiveness of our classification strategy will be demonstrated in the first instance  by studying a 
classical  symmetry-breaking channel  flow test problem and comparing results  
with classical stability results taken from the literature. 
Specifically, we compute approximations to  the velocity $\vfld{u}=[u_x,u_y]$ and 
pressure $p$ satisfying the Navier--Stokes  equations 
\begin{equation}\label{nse}
\rbl{\begin{aligned}
 \displaystyle  
{\partial {\vec u} \over \partial t}  + {\vec u}\cdot \nabla {\vec u}
-  {1\over \Re}\,\nabla^2 {\vec u} + \nabla p  &=  {\vec 0} , \\
  \nabla \cdot {\vec u} &=0 ,
  \end{aligned}}
\end{equation}
 in a rectangular duct  with a symmetric expansion.
 Details of the domain and a typical steady-state solution (below the critical Reynolds number 
 with an unperturbed inflow) are shown in Figure~\ref{step-figure-I}. 
 To be consistent with previous work, we will  characterise the flow Reynolds number
 $\Re= LU/ \nu$ in terms of the kinematic viscosity $\nu$, the maximum speed  at the inflow $U$ and
the height of the channel at the outflow $L$. 

 The associated steady-state  boundary conditions are 
\begin{itemize}  \itemindent -.3in \itemsep 3pt
\item 
parabolic  flow 
$\vfld{u}(-1,y) = (1-4y^2,0)$ at the inflow boundary 
$(-1,y),\,|y|\le 0.5$
\item
natural conditions $\nu \frac{\partial u_x}{\partial x}=p$, 
$\frac{\partial u_y}{\partial x}=0$  at the outflow boundary  $(16,y),\,|y|\le 1$
\item
no-flow conditions $\vfld{u}=0$  along the channel walls 
 $(x,\pm 1), 0\le x \le 16$; \\
  $(x, \pm 0.5), -1 \le x \le 0$;   $(0,y)$, $0.5 \le |y| \le 1$.
\end{itemize}
The position of the outflow boundary is far enough downstream that the flow is fully developed.
The key feature  of the base solution shown in Figure~\ref{step-figure-I} is that it is reflectionally
symmetric with respect to the centerline $y=0$, i.e., the stream function  $\psi$ satisfies 
 $\psi(x,y)=-\psi(x,-y)$. It follows that for the velocity, 
\begin{equation} \label{step-flow-character}
u_x(x,y)=u_x(x,-y), \quad u_y(x,y)=-u_y(x,-y).
\end{equation}
The symmetry of the  base solution can be confirmed  by observing that the two eddies 
downstream of the step  in Figure~\ref{step-figure-I} have the same reattachment length. 

The \ifiss\ software package~\cite{ifiss} is used to compute numerical solutions.
The spatial discretization is effected using  $Q_2$--$P_{-1}$  (biquadratic velocity; discontinuous linear pressure) 
mixed  approximation, see \cite[Section 3.3.1]{elman15}. 
Labelled data is generated by discretising the  flow domain into a uniform grid of 2112
square elements.\footnote{The open source MATLAB software toolbox  is described  in~\cite{elman14}. The 
grid is defined by setting the discretization  level  to  5.}
The dimension {$n_x$} of the resulting  system of ordinary differential equations is 23,842. 
The (unstabilised) TR--AB2 integrator described  in \cite[Sect.\,10.2.3]{elman15} is used for evolving the 
numerical solution. The time integration is started from quiescent flow $\vfld{u}_h=0$ in the channel domain. 
The initial time step is  {\tt 1e-9}. To model  the action of ``turning {on} the tap'' in a laboratory experiment,  the 
 inhomogeneous inflow boundary condition is smoothly ramped up. This is done by multiplying the 
 steady-state inflow boundary profile by the lifting function $1-e^{-10t}$. 
 
 The TR--AB2 integrator is run in fully nonlinear mode with a second-order linearisation step followed
 by two fixed-point iterations (a total of three ${n_x}\times {n_x}$ sparse matrix solves at each time step).
 The integrator is  run for a fixed number  ${n_t}$ of steps with the  time accuracy tolerance 
 set to {\tt 3e-6}. When generating  labelled data, ${n_t}$ is set to 1200. 
 
 \begin{figure}[!tbhp]
\centering
\vspace{2mm}
\includegraphics[width=.66\textwidth, height=.25\textwidth]{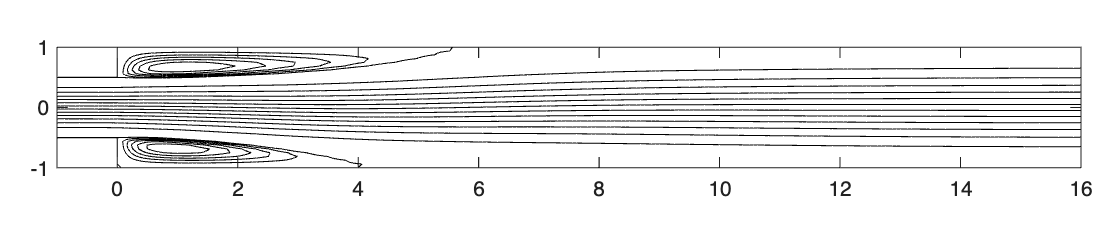}%
\vspace{-4mm}
\caption{Computed steady-state flow solution for $\Re=222.2$.} 
\label{step-figure-II}

\end{figure}

The specific  flow problem is referred to as a 1:2 expansion in the computational study described in \cite{drikakis97}. 
The flow exhibits a pitchfork bifurcation if the Reynolds number is increased  
past a critical value $\Re^*$ of approximately 220.  
At the critical value the rightmost eigenvalue of the linearised Jacobian 
matrix (see \cite{elman18}) changes from a real negative number (indicating linear stability) to 
a real positive number (instability). Just above the critical Reynolds number there are three steady
solutions: the symmetric flow solution is unstable and there are two stable asymmetric solutions with
the two recirculating eddies having a different length. One such solution computed at a Reynolds 
number $\Re>\Re^*$ is shown in  Figure~\ref{step-figure-II}. The second steady state  is the mirror 
image:  the top and bottom eddies  in  Figure~\ref{step-figure-II} are inverted.

\begin{figure}[!thp]
\centering
\includegraphics[width = 0.9\textwidth]{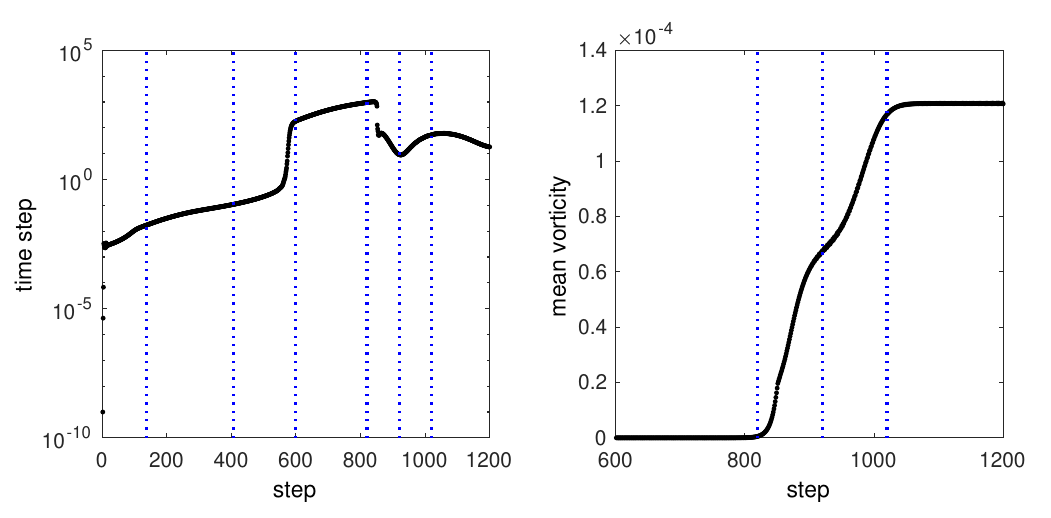}%
\caption{Time step history  and mean vorticity evolution for $\Re=222.2$. The vertical lines 
refer to specific time steps that are identified in the text.}
\label{step-figure-III}
\end{figure}

The  {\it mean vorticity}  $\omega(t)$, or the average vertical velocity at the outflow ($\partial\Omega_N$)
 \begin{equation} \label{meanvorticity}
 \omega(t) =  \int_\Omega \nabla \times \vfld{u}(\cdot,t) \,  \rm{d}\Omega
 =  \int_{\partial\Omega_N} \!\!\!\! u_y(\cdot,t) \, \rm{d}s  ,
 \end{equation}
provides a convenient way of assessing the degree of departure from 
the  steady symmetric  flow (for which $\omega$=0). 
The evolution of the time step when computing the flow solution in  Figure~\ref{step-figure-II} is
shown in Figure~\ref{step-figure-III}. Following the physics and checking snapshots of 
the evolving flow solution one can identify four distinct phases in the time integration process:
\begin{itemize}  \itemindent 0pt  \itemsep 3pt
\item 
The first phase is the propagation of the inlet pulse from the inflow  to the channel 
expansion, that is,  the first time unit of integration. This milestone
is reached after 136 time steps, at which point the  {step size} is  0.017 and the
two downstream eddies are just starting to form.
\item 
The second stage is the propagation of the inlet pulse from the channel expansion to
the outlet. This milestone is reached after  {a further 270 time steps}.
At this point in time a symmetric downstream eddy profile 
is fully established. Following this,  the  time step  grows rapidly (by two orders of magnitude) as the flow
solution relaxes---an (unstable) symmetric eddy profile (similar to that {in} Figure~\ref{step-figure-I}) 
can be  observed at {step number} 600.
\item
The third phase is the transition from the symmetric eddy profile to an
asymmetric profile. After 820 steps the  mean vorticity rapidly increases by two orders of 
magnitude.  An asymmetric eddy profile is clearly formed at {step number} 920.
\item
The final phase is a relaxation into the final asymmetric eddy profile shown 
 in Figure~\ref{step-figure-II}. The mean vorticity stops increasing at step 1020
  before it settles to a steady value of $\omega=1.2\times 10^{-4}$.
There is nothing to be gained by continuing the time integration beyond this point. Once 
the symmetry is broken there is no way back.
\end{itemize}

From a machine-learning perspective the mechanics of assigning a label to a computed 
flow solution needs to be definitive. The flow solution in Figure~\ref{step-figure-I} is 
computationally symmetric because the $x$ coordinate  where the upper eddy reattaches on the top
boundary is close enough to the $x$ coordinate  where the lower eddy reattaches on the bottom
boundary. The asymmetry of the flow in  Figure~\ref{step-figure-II} is self-evident if we compare the
 computed vorticity (wall shear stress) on the top and bottom boundaries, as shown in 
 Figure~\ref{step-figure-IV}.   The location $x>0$ where the wall shear stress is zero is the point where
  the eddy reattaches to the wall. These locations are obviously  different: the vorticity on the
  lower boundary changes sign between boundary grid nodes 65 and 66, whereas  the 
  vorticity on the upper  boundary changes sign between  grid nodes 88 and 89. This
  simple test classifies the flow in  Figure~\ref{step-figure-II} as ``bifurcated'' ($\ell^j=1$). 
  Applying the same test to the flow in  Figure~\ref{step-figure-I}   the indices where the  sign changes 
  are within one  grid point of each other---this classifies the flow  as being  symmetric ($\ell^j=0$).
  This is a very strict test of the flow symmetry---it  is  appropriate in cases where the
  loss of stability is a smooth function of the bifurcation parameter(s) and gives reliable results
 whenever the spatial resolution is sufficiently fine.\footnote{If
  the sign change  on the top was between  nodes 74 and 75, and on the  bottom was between nodes 
 73 and 74  or  between 75 and 76 we would still  classify the flow  as being symmetric ($\ell^j=0$).}
Further justification  will be provided later.
 
 \begin{figure}[!thp]
\centering
\includegraphics[width = 0.7\textwidth]{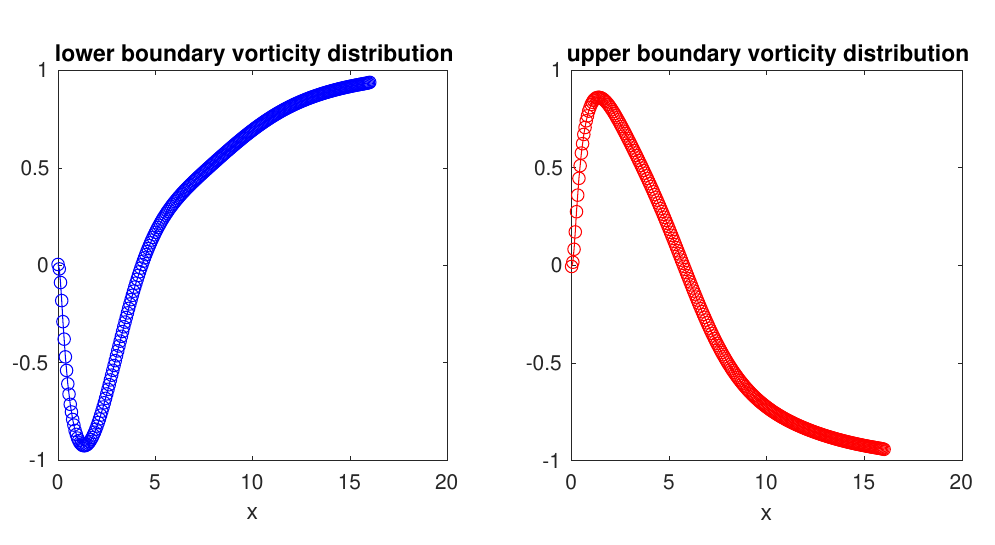}%
\caption{Computed wall shear stress  for $\Re=222.2$.} 
\label{step-figure-IV}
\end{figure}

As noted earlier, classical eigenvalue analysis provides 
no information about the size of  asymmetric perturbation needed \rbl{to  initiate}  a  transition to an asymmetric
flow profile from an unstable symmetric flow pattern. Herein,  to replicate a practical laboratory
experiment we will assess  the
symmetry or otherwise of an evolving flow  generated with a {\sl distorted} steady-state parabolic inlet profile,
\begin{equation} \label{inflow}
{  u_x (-1,y) = \delta \cdot \sin( 2\pi y) + (1-4y^2 )},  \end{equation}
where the parameter $\delta$ measures the size of the perturbation. The
skewed inlet profile effectively disconnects the pitchfork
bifurcation structure---the only asymmetric solution that is computed when 
 $\delta$ is greater than unit roundoff ($\varepsilon$) is  a flow profile with
 a longer eddy at the bottom compared to the top. 
 
\subsection{Informed data test results} \label{sec:informed}
Making use of a priori knowledge, labelled data was generated for two different sets of parameters  $\lambda=(\Re,\delta)$. 
 The first set of results is generated with 35 uniformly spaced Reynolds numbers between 209.6 and 225.7 with 10
logarithmically spaced perturbations between $10^{-16}$ and $10^{-3}$. 
The second set is generated with 30 uniformly spaced Reynolds numbers between 168.1 and 222.2 and 
5 logarithmically spaced perturbations between $10^{-5}$ and $5\times 10^{-2}$.  Combining these two data
sets into a single set $\SS_1$, 281 data points  were labelled as symmetric flows  and 219 were classed as 
asymmetric. 

\begin{figure}[!thp]
\centering
\includegraphics[width = 0.99\textwidth]{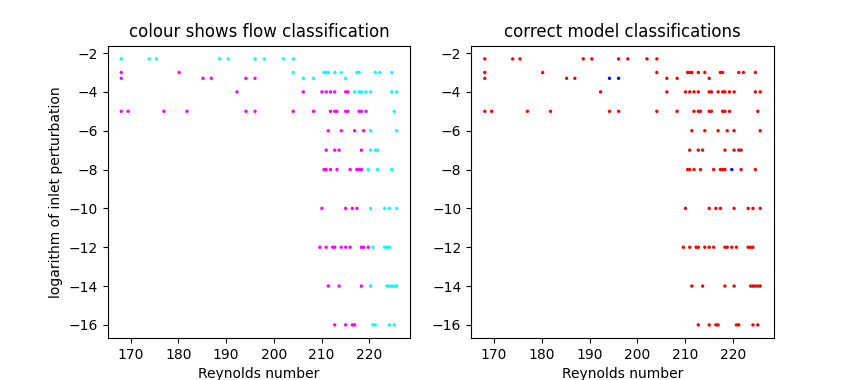}%
\caption{Symmetric step channel flow: typical test data classification (left) and prediction accuracy  (right).
Red dots on the right are correct predictions,  blue dots are incorrect.} 
\label{step-figure-V}
\end{figure}

To \rbl{investigate} the predictive capability of the surrogate model, the  data set  $\SS_1$ is randomly divided  into a 
training set with 375  data points and a test set with 125 data points.  
The classification of the test data in one such division is shown in  Figure~\ref{step-figure-V}. 
In this representative data split the surrogate model generated by  the training data
correctly classifies 122 of the test data points with 2 asymmetric flows incorrectly 
classified as being symmetric and one flow incorrectly classified as being unsymmetric.
The \rbl{data splitting} was repeated multiple times. The mean precision after 5 independent  \rbl{
generalisation tests of this type was 0.964. The  mean F1 score for the 5  tests was 0.969.} 

\rbl{The data-splitting  test  results are unsurprising. It is clear} that 
simpler strategies such as a $k$ nearest neighbour or a shallow decision tree classifier would be equally effective 
ways of  visualising the decision boundary in cases like this where one is 
prepared to  generate results with a rich set of well-chosen  training points. 
The attractive features of the shallow neural network classifier  lie in its
{\it simplicity} (it is essentially parameter-free: the architecture is fixed and the optimisation strategy
does not need to be specially tuned),  its {\it practicality} (the raw output of the network is the probability
of the bifurcation being observed), \rbl{its {\it generalisability} (to high-dimensional parameter vectors),}
and its {\it speed} (the total  time taken to complete  the  training  
was less than 4 minutes when  the open-source Python code was run on a 2018 MacBook Pro).

\begin{figure}[!thp]
\centering
\includegraphics[width = 0.85\textwidth]{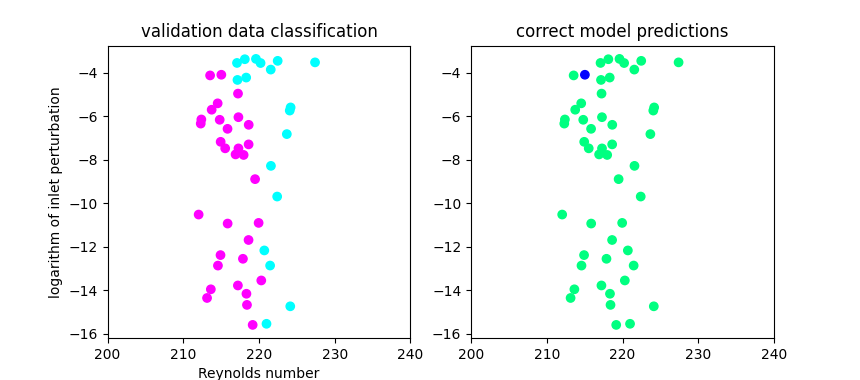}%
\caption{Symmetric step channel flow: refined grid classification (left) and prediction accuracy  (right).
Green dots on the right are correct predictions,  blue dot is incorrect.} 
\label{step-figure-VII}
\end{figure}

To assess the sensitivity to discretisation parameters and validate \rbl{the surrogate model generated by 
the  500 data points,}  a second set of 50 classification data points were randomly generated
 close to the classification boundary with a refined spatial grid (level 6 in \ifiss). In these calculations, 
 the dimension ${n_x}$ of the algebraic system is  increased from 23,842 to 94,146 
and  ${n_t}$, the number of time steps, is increased from 1200 to 1600. 
The results are shown in  Figure~\ref{step-figure-VII}.  \rbl{It is reassuring to observe that only one  parameter 
combination was misclassified by the surrogate model.  The results in the next section demonstrate 
that data points with perturbations of magnitude between $10^{-3}$   and $10^{-4}$ prove to be the
most difficult to classify correctly.}
  
 \subsection{Blind data test results} \label{sec:blind}
  As noted above, there is \rbl{a degree} of uncertainty in the classification  
 of flows with perturbations  between \rbl{$10^{-3}$   and  $10^{-4}$}. To investigate this,  our bifurcation analysis was restarted 
 from scratch with a \rbl{coarse parameter set $\SS_2$ of 105 data points, generated by taking 7 uniformly spaced values of $\Re$ 
between 160 and 240 and 15 logarithmically spaced values  of  $\delta$ between  $10^{-5}$  and $10^{-2}$}.
The visualisation of  the bifurcation boundary  in the \rbl{left plot} in Figure~\ref{step-figure-VI} is generated  
by training the neural network on the set of 105 data points, sampling 
the parameter set at 200,000 points uniformly distributed in the region of interest and 
\rbl{colouring the probability of observing a bifurcated solution at these parameter values using a cyclic colormap.}
The uniformly  sharp transition is the striking feature of this picture.

 \begin{figure}[!thp]
\centering
\includegraphics[width = 0.95\textwidth]{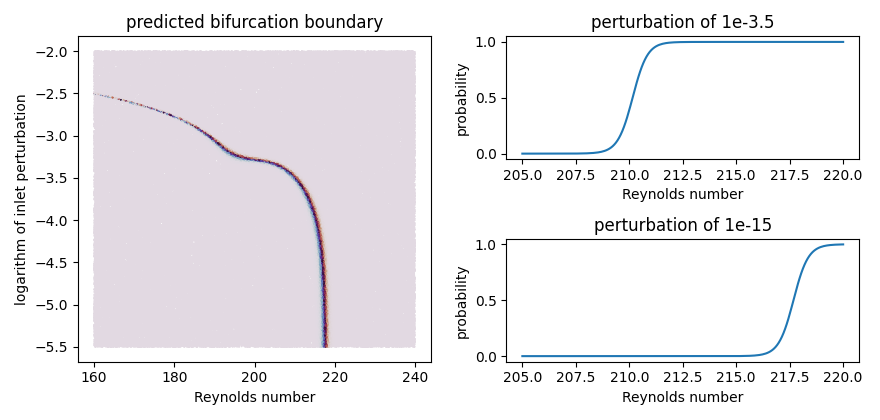}%
\caption{\rbl{Symmetric step channel flow showing surrogate predictions  from 
trained neural network using 105 uniformly spaced data points. Colour variation highlights the bifurcation boundary 
solution (left plot). Estimated probability of observing a bifurcated solution for a fixed perturbation 
 (right plot).}} 
\label{step-figure-VI}
\end{figure}

\rbl{ The two plots on the right of Figure~\ref{step-figure-VI} show the estimated probability  of observing a bifurcated solution 
for perturbations of fixed magnitude. These probability plots were generated by uniformly sampling   the  $\SS_2$  
surrogate network  at 200 values of $\Re$ with fixed values, $\delta=10^{-3.5}$ and  $\delta=10^{-15}$.}

\rbl{An important observation is that the  probability distribution generated by the blind dataset surrogate 
 is perfectly  consistent with estimates of  the critical Reynolds number $\Re_*$ for unperturbed flow ($\delta=0$) 
 that have been published in the literature.
Results for three studies reported in Table~II in~\cite{drikakis97}  suggest that $\Re_*$ is between 216 and 219, but 
the authors also remark  that ``the exact value of the critical Reynolds number is difficult to fix
because it depends on grid resolution, but it is close to 216.''  This remark  is supported by the
eigenvalue results  in~\cite[Fig.\,4]{elman18} showing that computed estimates of the  critical  
eigenvalue governing linear stability move to the left  as the  spatial resolution is increased.}

\subsection{Validation of the blind test results} \label{sec:valid}
\rbl{To confirm the predictive capability of the coarse grid surrogate model,  a refined  data set  $\SS_3$ was
generated with 1476 points, consisting of 41 uniformly spaced points in $\Re$ between  160 and 240 and 
 36  logarithmically spaced values  of  $\delta$ between  $10^{-5}$  and $10^{-2}$.  The
resulting predictions of the bifurcation boundary, and estimates of observing a bifurcated solution for a fixed perturbation 
are shown in Figure~\ref{step-figure-VII}. When the coarse grid sample point neural network surrogate  is used to predict the 
label at the fine grid sample points the  precision was 0.971 and the F1 score was 0.970
(1457 labels were correctly predicted). }

 \begin{figure}[!thp]
\centering
\includegraphics[width = 0.95\textwidth]{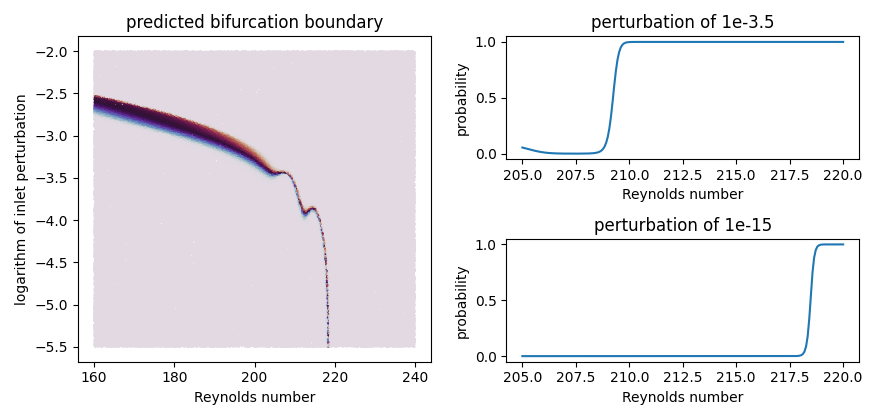}%
\caption{\rbl{Symmetric step channel flow showing surrogate predictions  from 
trained neural network using 1476 uniformly spaced data points. Colour variation highlights the bifurcation boundary 
solution (left plot). Estimated probability of observing a bifurcated solution for a fixed perturbation 
 (right plot).}} 
\label{step-figure-VII}
\end{figure}

\rbl{
Comparing the results in Figures~\ref{step-figure-VI}  and~\ref{step-figure-VII}
we note four distinctive features:
\begin{itemize}  \itemindent 0pt  \itemsep 3pt
\item
The zone of uncertainty  in the predicted  bifurcation boundary with $\delta <  10^{-3.5}$  is 
significantly reduced  in the fine grid surrogate (no surprise, there are more sample points in
the uncertain  region).
\item
The fine grid surrogate introduced more uncertainty in the position of the boundary for 
perturbations $\delta$ larger than  $10^{-3.5}$. 
\item
The coarse grid surrogate gives an excellent prediction of the probability of 
observing a bifurcation for the two parameter values that are highlighted.
\item 
With  a visually imperceptible inlet profile  perturbation  of 
magnitude  of  $10^{-3.5}$  the critical Reynolds
number  decreases from the unperturbed value of  220 to  a value close to  210. This is a striking and surprising result.
\end{itemize}}


\section{Classification of instabilities in  thermal convection} \label{sec:application}
The broad applicability of our  classification strategy will be demonstrated by exploring
the bifurcation structure of two representative problems
that arise when modelling buoyancy driven flow.
The first case we consider  is a 
symmetry-breaking pitchfork bifurcation and the second is a Hopf bifurcation from a steady
flow solution to a periodic flow. We note that  both problems are members of the
 set  of  canonical bifurcation test problems discussed by Wubs \& Dijkstra  in~\cite[Chap.\,10]{wubs23}. 
\rbl{The PDE system that we use to model such flows is  given by}
\rbl{
\begin{equation} \label{boussinesq}
\begin{aligned}
{\partial {\vec u} \over \partial t}  + {\vec u}\cdot \nabla {\vec u}
-  {\nu} \,\nabla^2 {\vec u} + \nabla p  &=  \theta {\vec j},  \\
  \nabla \cdot {\vec u} &=0 ,   \\
 {\partial \theta \over \partial t}  + {\vec u}\cdot \nabla \theta
-  {\epsilon} \,\nabla^2 \theta  &=  0  ,
  \end{aligned}
  \end{equation} }
\rbl{where $\theta$ is the temperature, and we observe that the Navier--Stokes equations (\ref{nse}) are  
being driven by the  buoyancy forcing term $ \theta {\vec j}$ acting downwards. The fluid velocity ${\vec u}$ 
is always set to zero on the  domain boundary. Working in 
dimensionless units the diffusion parameters $\nu$ and $\epsilon$  in (\ref{boussinesq})  can be 
expressed in terms of  the characteristic Rayleigh number  \Ra\ and Prandtl number Pr via}
\begin{equation}
\rbl{
\begin{aligned}
\nu = \sqrt{{\hbox{Pr}\over \Ra}}, &\quad \epsilon = {1\over \sqrt{\hbox{Pr} \Ra}}.
\end{aligned}}
\end{equation}
\rbl{Note that the two diffusion parameters are close to each other when Pr is close to 1; for
example, when modelling the thermal convection of air, but are significantly different 
for fluids  with Pr $\gg 1$.}

 \subsection{A Rayleigh--B{\' e}nard convection problem} \label{sec:benard}
Following their discovery by B{\' e}nard in the laboratory in 1900, the development of recirculating convection 
cells has been extensively   studied: both  theoretically, see~\cite[Chap.\,6]{drazin02}, and
computationally, see Gelfgat~\cite{gelfgat99a}. A state-of the-art algorithm that combines 
Newton's method with a simple deflation strategy  can be found in Boulle et al.~\cite{boulle22}. 

The  focus here will be on a  geometric configuration  that is outside the 
realm of  Rayleigh's shallow layer  linearised stability analysis. Specifically we consider a
(tall) 1:2 aspect ratio cavity with insulated vertical boundaries  $x=0$ and $x=1$.
The lower boundary $y=0$ is hot and the top boundary $y=2$ is cold. If
the imposed temperature difference is sufficiently small  then there is no motion and the 
buoyancy effect of rising fluid is balanced by the effect of gravity acting downward. If
the temperature difference is increased beyond a critical value  then the 
symmetry of the solution is broken and a steady recirculation is established that
moves  in a clockwise (upper branch) or an anticlockwise {(lower branch) direction}. 

\begin{figure}[!thp]
\centering
\includegraphics[width = 0.6\textwidth]{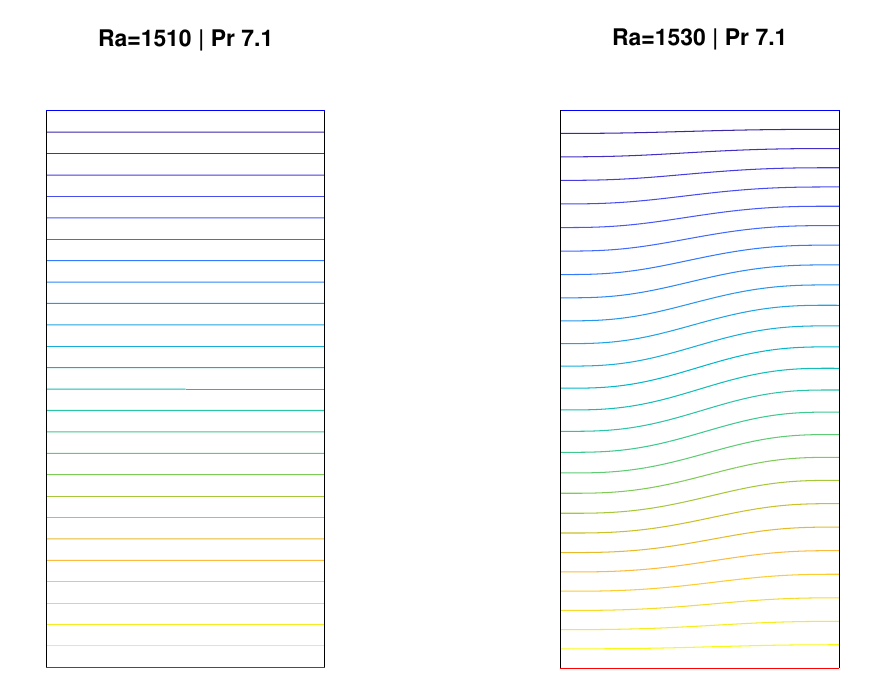}%
\caption{Computed isotherms for unperturbed  Rayleigh--B{\' e}nard convection: {steady conducting state} (left)
and bifurcated solution (right).}
\label{benard-figure-|}
\end{figure}

The fluid modelled is water so the Prandtl number 
{Pr = $\nu/\epsilon$, where $\nu$  is the kinematic viscosity and $\epsilon$ is the thermal diffusivity}, 
 is set to  7.1.  As in the previous section, we set up the bifurcation  problem $\lambda$ having  
two parameters,  \Ra\ and  $\delta$. The first of the parameters  is the Rayleigh  number that is  associated with
the standard {nondimensionalization  of the governing (Boussinesq) system, defined here to be}
\begin{equation*}
{
\hbox{\Ra} = {g\beta(T_{\hbox{hot}} - T_{\hbox{cold}}) d^3 \over
\nu \epsilon},
}
\end{equation*} 
{where $g$ is the acceleration due to gravity, $\beta$ is the thermal expansion coefficient of the fluid
and $d$ is the distance between the hot and cold boundaries.}
{The  Rayleigh  number} conveniently characterises 
the degree of instability of the system. Computed steady-state temperature profiles for two values are illustrated 
in  Figure~\ref{benard-figure-|}.  We observe that when  \Ra\ is 1510 the no-flow {conducting state}
solution is stable, whereas when  \Ra\ is  increased to 1530  it is unstable.
The asymmetric solution shown in the right plot corresponds to an anticlockwise circulation. The  other
stable steady state when $\hbox{\Ra}  = 1530$  features an  clockwise circulation of the same magnitude
 with the isotherms reflected in the line of symmetry $x=0.5$.
 Note that, in this case, the  {\it kinetic energy}  
 \begin{equation} \label{kinetic}
k(t) =  {1\over 2} \int_\Omega \vfld{u}(\cdot,t)  \cdot  \vfld{u}(\cdot,t)  \,  \rm{d}\Omega,
\end{equation}
provides a convenient way of assessing the degree of departure from the  equilibrium flow. 

The second parameter within the vector $\lambda$ represents the magnitude
of an asymmetric perturbation of the constant temperature at the base of the cavity. 
In addition, to model  the action of ``heating the pan'' in a laboratory setting,
the temperature variation on the horizontal walls {is} asymmetrically ramped up
in time as was done  in section~\ref{sec:validation}. The resulting boundary
conditions are
\begin{align} \label{hotwalls}
\begin{aligned}
T(\vfld{x},t) &= \left \{ {1\over 2} + \delta \cdot \sin(2\pi x) \right\} (1-e^{-10t}),  &  0\leq x\leq 1, y=0 , t>0,\\
T(\vfld{x},t) &= -{1\over 2}(1-e^{-10t}) ,  &  0\leq x\leq 1, y=2, t>0.
\end{aligned}
\end{align}

Numerical solutions are computed with the default  \ifiss\  spatial approximation 
 $Q_2$--$Q_1$--$Q_2$ of the Boussinesq flow equations (biquadratic temperature)
  \cite[Sect.\,11.3]{elman15}.  Labelled data is generated by discretising the  cavity domain 
into a uniform grid of  $32\times 64$ square elements.
The dimension ${n_x}$ of the resulting algebraic system in this case is 27,300. The linearised TR--AB2 
integrator described  in \cite[Sect.\,11.2]{elman15}  (one ${n_x}\times {n_x}$  sparse matrix solve per time step)
{is} used  for generating the numerical solution. The linear equation systems 
are near singular (in exact arithmetic {there is} a constant pressure vector in the null space). 
Numerically singular systems that arise in the course of  the time integration are dealt with
 by constructing and solving an augmented (regularised) system, see Bochev \&
 Lehoucq~\cite{bochev05}. 
The initial time step is  {\tt 1e-9}. The stabilised TR--AB2 integrator is  run for a fixed number 
 ${n_t}$ {steps} with the  time accuracy tolerance  parameter set to {\tt 1e-6}. 
When generating  labelled data, ${n_t}$ is set to 800. 
 
\begin{figure}[!thp]
\centering
\includegraphics[width = 0.9\textwidth]{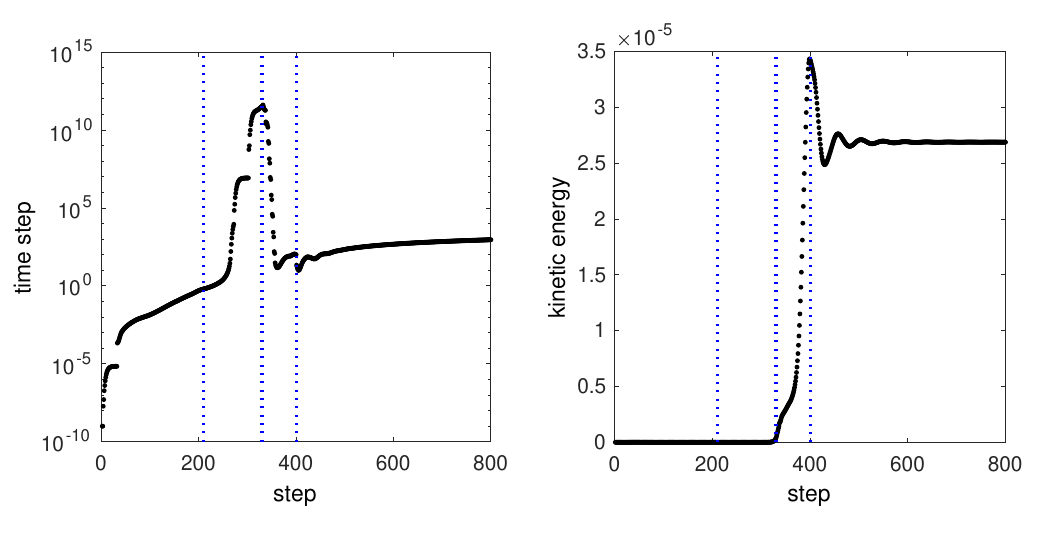}%
\caption{Step  size history (left) and kinetic energy (right) for $\Ra=1530$, $\delta=0$.
The vertical lines refer to specific time steps that are highlighted in the text.}
\label{benard-figure-|I}
\end{figure}

The time step evolution when computing the bifurcated solution 
in  Figure~\ref{benard-figure-|} is shown in Figure~\ref{benard-figure-|I}. Following the physics and 
checking snapshots of  the evolving temperature solution one can again identify four distinct phases
in the time integration process:
\begin{itemize}  \itemindent 0pt  \itemsep 2pt
\item 
The first phase is the development of the equilibrium solution as the horizontal walls are heated/cooled
from zero to their final values of $\pm 0.5$.\footnote{The visible  jumps in the time step after 30 
and after 270 steps correspond to an TR--AB2 averaging step. If averaging is not done  then 
the time step stagnates before  the equilibrium solution is established. Averaging was done 
every 30 steps until step 300, but was not invoked for the final 500 steps.}
A  tiny jump in the kinetic energy between  time steps 33 and  66  (too small to feature
on the plot) is associated with the motion of the fluid needed to initiate  
the transition to the equilibrium state. 
A linear temperature profile  is clearly established after  20 time units and 210 time steps.
\item 
 After this  milestone,  the  step size  grows rapidly (by twelve orders of magnitude) as the temperature
solution relaxes---an (unstable) {\it symmetric} temperature profile (as in the left plot in 
Figure~\ref{benard-figure-|}) can be  observed at time step 320.
\item
The third phase is the transition from the symmetric profile to an
asymmetric profile. After 330 steps the adaptive integrator starts rejecting steps and the step size
is reduced by eleven orders of magnitude over the next 40 time steps. This decrease in
step size is associated with a rapidly increase in the kinetic energy. An asymmetric 
temperature profile corresponding to {an} anticlockwise recirculation is clearly formed at time 
step 400. In contrast, when generating the solution with \Ra=1510 shown in Figure~\ref{benard-figure-|},
there is no increase  in the kinetic energy after 300 steps and no associated reduction in the step size.
\item
The final phase is the relaxation to the temperature profile illustrated
 in the right plot in Figure~\ref{benard-figure-|}. The kinetic energy
 ``overshoots'' before being quickly damped to a perfectly steady asymmetric profile.  
 As noted in  section~\ref{sec:validation}, there is nothing  to be gained by continuing the  
 integration beyond this point.
\end{itemize}

\begin{figure}[!thp]
\centering
\includegraphics[width = 0.95\textwidth]{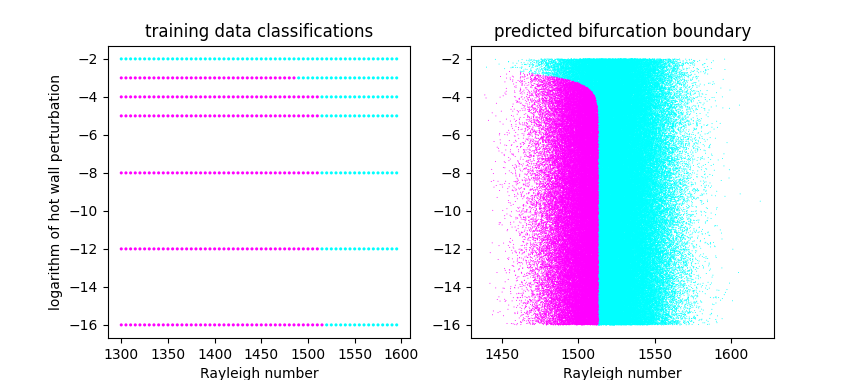}%
\caption{Rayleigh--B{\' e}nard convection: data classification (left);  surrogate predictions  from 
trained neural network  (right).
Red dots correspond to $\ell^j$=0 and blue dots correspond to $\ell^j$=1.} 
\label{benard-figure-III}
\end{figure}

\begin{figure}[!thp]
\centering
\includegraphics[width = 0.9\textwidth]{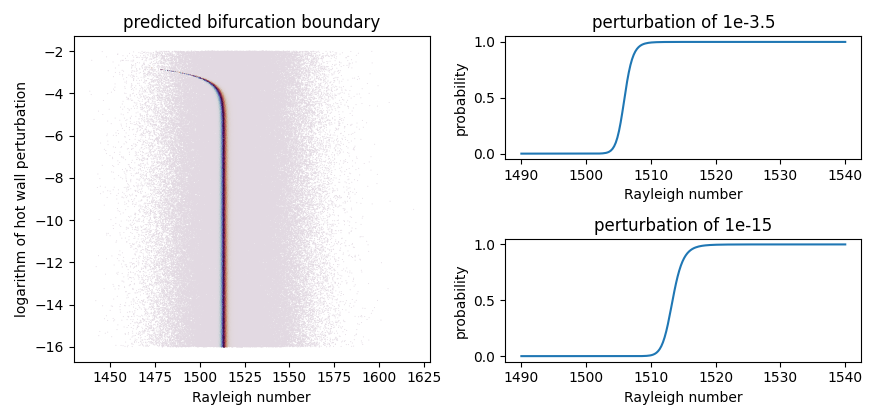}%
\caption{\rbl{Rayleigh--B{\' e}nard convection  showing surrogate predictions  from 
trained neural network using 420 data points. Colour variation highlights the bifurcation boundary 
solution (left plot). Estimated probability of observing a bifurcated solution for a fixed perturbation 
 (right plot).}} 
\label{benard-figure-IV}
\end{figure}

The dataset  $\SS$  was generated by taking 60 uniformly spaced Rayleigh numbers between 1300 and 1600 and 7
logarithmically spaced perturbations between $10^{-16}$ and $10^{-2}$.   A flow problem is classified 
as ``bifurcated'' $(\ell^j=1)$ if the final  kinetic energy  was greater than $2\times10^{-7}$ (this is clearly the case
for the parameter combination in Figure~\ref{benard-figure-|I}). With this test in place,
 254  of the data points $\SS$  were labelled as a system in the conducting state  and 166 were classed as 
asymmetric. The precise classification is shown in Figure~\ref{benard-figure-III}.  Sampling the trained 
neural network at 200,000 points leads to  the prediction of the  bifurcation boundary shown in the right plot. 

\rbl{The visualisation of the predicted bifurcation boundary in  Figure~\ref{benard-figure-IV} follows the strategy
described in section~\ref{sec:blind}. As in that case, the uniformly  sharp transition is the prominent feature.
From this set of labelled data, one can be confident that the critical Rayleigh number for perfect uniform heating of the cavity bottom
is in the interval $(1510,1520)$.  In this case the visually imperceptible temperature profile  perturbation  of 
magnitude  of  $10^{-3.5}$  decreases the critical Rayleigh number  from the unperturbed value 
to  a value close to 1505. Comparing the fixed perturbation probability profiles with those in 
Figure~\ref{step-figure-VII}, we see that the effect of the perturbation is similar in both cases.
This is probably to be expected.}

\subsection{A differentially heated cavity problem} \label{sec:wubs}
The second problem is motivated by laboratory experiments originally  performed by Elder~\cite{elder65}.
The experiments were performed over 50 years ago by
observing the motion of aluminium powder suspended in viscous oil with a Prandtl number 1000. 
The aim here is to extend our understanding  with  a view to assessing  the stability of the laboratory
configuration for  a variety of  fluids with Pr  ranging from 7.1 (water) to 1000 (glycerol). We note that the bifurcation 
structure  for fluids with low Prandtl numbers has also been  extensively studied 
over the past 25 years, see Gelfgat et al.~\cite{gelfgat99b}. Elder's  original experimental set up 
can be modelled as  a (tall) 0.051:1 aspect ratio  cavity with insulated horizontal boundaries  $y=0$ 
and $y=1$. The left boundary $x=0$ is hot and the  right boundary $x=0.051$ is cold. If the imposed 
temperature difference is sufficiently small  then a steady flow solution will be established with a 
clockwise primary flow recirculation. 

Numerical solutions are  computed with $Q_2$--$Q_1$--$Q_2$  spatial approximation
using \ifiss. Labelled data is generated by  discretising the cavity domain into a nonuniform grid of 
 $32\times 256$ rectangular elements
that is geometrically stretched away from the walls.\footnote{The 
grid is defined by setting  the horizontal stretch factor to 2  and the vertical stretch factor to 1.5.}
 The dimension ${n_x}$ of the resulting algebraic system in this case is 108,516. 
 The TR--AB2 integrator is run in fully nonlinear mode with a second-order linearisation step followed
 by two fixed-point iterations as in section~\ref{sec:validation}.
 The integrator is  run for a fixed number  ${n_t}$ {steps} with the  time-accuracy tolerance parameter
 set to {\tt 2e-5}. When generating  labelled data ${n_t}$ is set to 2000. 
 To model  the action of heating the walls  in a laboratory setting
the temperature variation on the vertical walls {is} ramped up in time, multiplying the 
steady-state constant boundary temperature profile  by the lifting function $1-e^{-10t}$. 

If the Rayleigh number  is above a critical  value 
then the steady solution loses stability  via a Hopf bifurcation. 
The associated Rayleigh number that was  identified from  the laboratory experiments  is 
$2.26\times 10^9 \pm 30\%$\footnote{The definition of the Rayleigh number
 in~\cite{elder65} is scaled by a factor of ${d}^3$.}, {the} large uncertainty {is} attributed to the 
 difficulty of detecting the onset of a very weak secondary flow. 
A precise critical value  of  $2.88\times 10^9$ is identified  in~\cite[Sect.\,10.4]{wubs23}
by carefully tracking eigenvalues. Computational results for 
 a Rayleigh number  just above this value   typically evolve  to a periodic flow solution
  with small  ``ripples''  along the vertical centreline $y=0.5$. 
Such an instability  is apparent in  the volume averaged kinetic energy history 
for  a Rayleigh number of  $2.89\times 10^9$  that is presented  in Figure~\ref{wubs-figure-|},
wherein a stable oscillation with a period of 683 time units can be clearly identified after the first 1000 steps
of integration. In contrast, when the computed kinetic energy is plotted  on the same scale for
a slightly smaller Rayleigh number, say Ra = $2.87\times 10^9$,  there are no  oscillations 
visible after 1000 time steps, {see  Figure~\ref{wubs-figure-|V}.}

\begin{figure}[!thp]
\centering
\includegraphics[width = 0.8\textwidth]{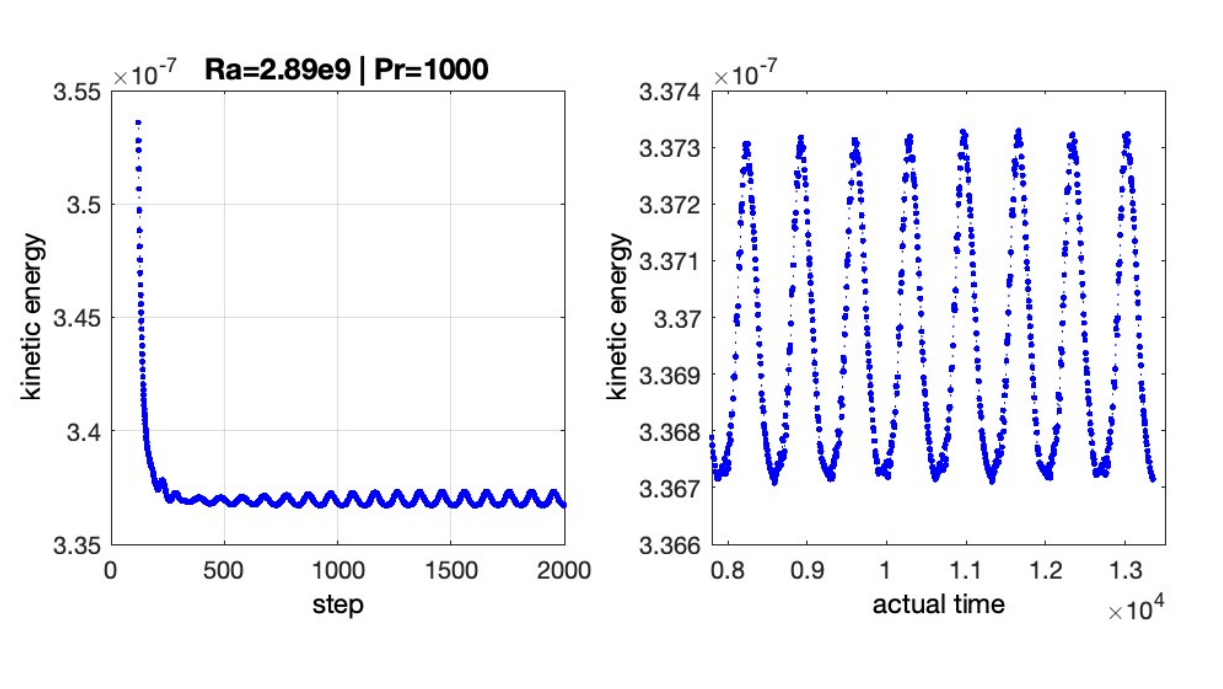}%
\caption{Evolution of the kinetic energy for a  differentially heated cavity with \Ra=$2.89 \times 10^9$ 
and Pr=1000: time steps  120--2000 (left), time steps 1200--2000 (right).} 
\label{wubs-figure-|}
\end{figure}

\begin{figure}[!thp]
\centering
\includegraphics[width = 0.8\textwidth]{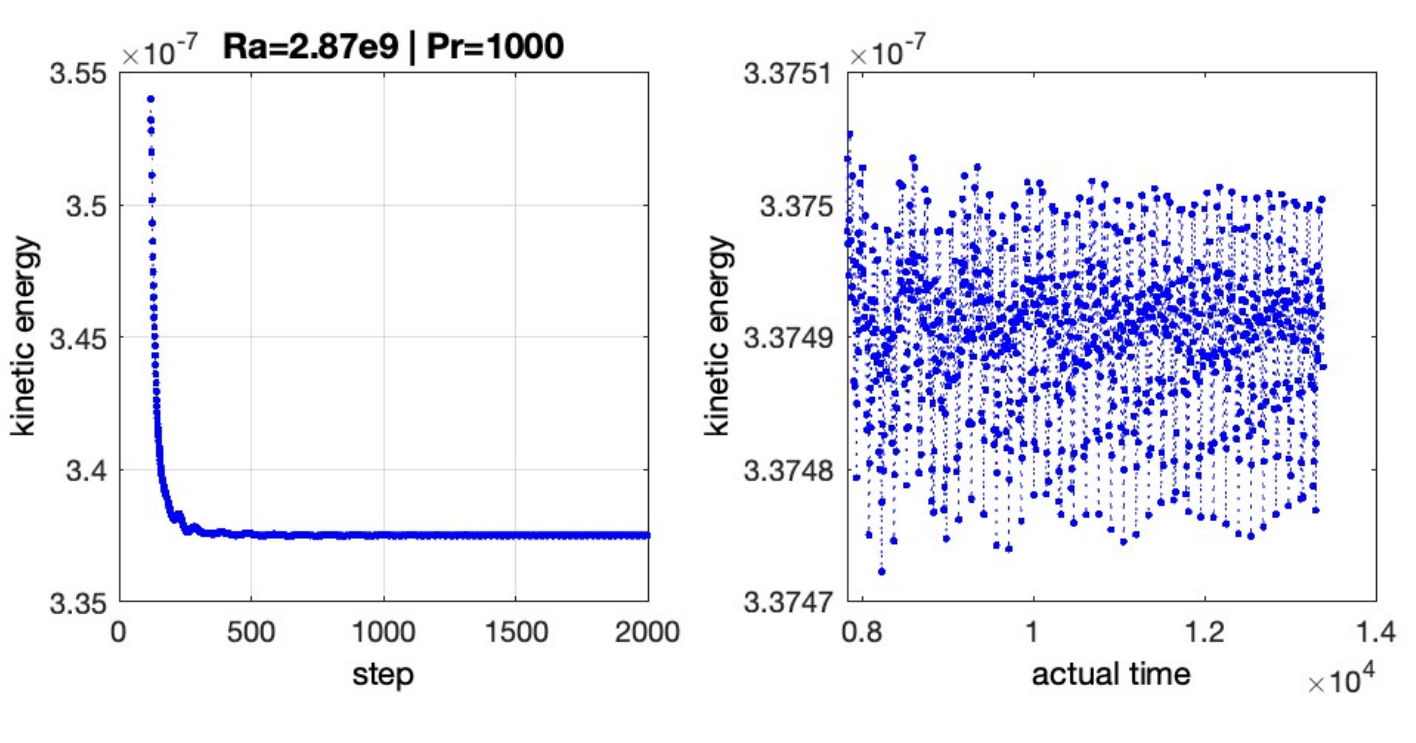}%
\caption{Evolution of the kinetic energy for a  differentially heated cavity with \Ra=$2.87 \times 10^9$ 
and Pr=1000: time steps  120--2000 (left), time steps 1200--2000 (right).} 
\label{wubs-figure-|V}
\end{figure}

There are only two distinct phases in the evolution of the time step when using stabilised TR--AB2 to  
compute the bifurcated solution in  Figure~\ref{wubs-figure-|}:
\begin{itemize}  \itemindent 0pt  \itemsep 3pt
\item 
The first phase is the development of the primary recirculation as the vertical walls are heated/cooled
from zero to their final values of $\pm 0.5$.  There is no analytical solution in this case---even a small temperature 
difference will generate a nontrivial flow in the cavity. The primary recirculation is fully developed after
100 time steps, which translates to about 200 units of  time.
\item
The second phase is the development of the oscillating solution. After the first 350 steps the
step size variation is what one would expect when following the physics of a periodic solution.
\end{itemize}

\begin{figure}[!thp]
\centering
\includegraphics[width = 0.95\textwidth]{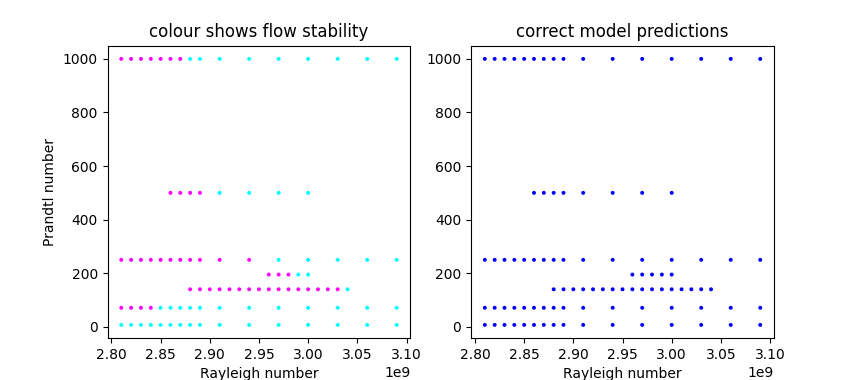}%
\caption{Unsteady convection: training data classification (left) and prediction accuracy  (right).
Blue dots on the right are correct predictions.} 
\label{wubs-figure-II}
\end{figure}


\begin{figure}[!thp]
\centering
\includegraphics[width = 0.95\textwidth]{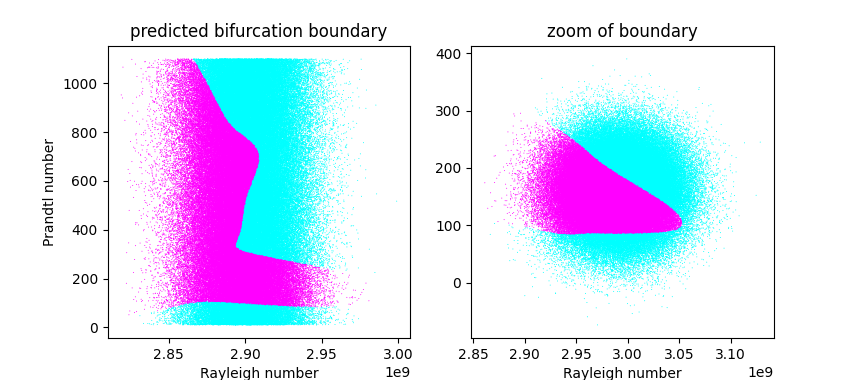}%
\caption{Unsteady convection: surrogate predictions  from the
trained neural network sampled  in the neighbourhood of the classification boundary.}
\label{wubs-figure-III}
\end{figure}

 The parameters defining the bifurcation  problem that is the focus here  are the 
Rayleigh number (\Ra) and the Prandtl number (Pr). The dataset  $\SS$  was generated for 7 specific
values of Pr; namely, 7.1, 71, 140, 195, 250, 500 and 1000.   We computed results for  different 
values of \Ra\  between $2.81 \times 10^9$ and $3.09 \times 10^9$. 
A flow problem is classified as ``bifurcated'' $(\ell^j=1)$ if, when looking 
at the final 800 time steps of the computation, the ratio of the difference between the maximum and  
the minimum energy $k$ divided by the mean value of $k$ is greater than  $6\times 10^{-4}$  (as is 
the case for the parameter combination in Figure~\ref{wubs-figure-|}). With this test in place,
 45  of the data points $\SS$  were labelled as a system that reached a steady state  and 49 points
 were classed as being oscillatory. The precise classification is shown in Figure~\ref{wubs-figure-II}. 
   
Sampling the trained  neural network at 200,000 points leads to  the prediction of the 
 bifurcation boundary that is shown in  Figure~\ref{wubs-figure-III}.
 The results illustrated have the following distinctive features:
 \begin{itemize}  \itemindent 0pt  \itemsep 3pt
\item 
Both the raw data for Pr=1000  and the bifurcation point  predicted by the 
trained neural network are in close agreement with the results in~\cite{wubs23}.
\item 
When the Prandtl number  is reduced from 1000 by an order of magnitude, the critical 
Rayleigh number increases, eventually reaching a maximum  value of   
$3.05 \times 10^9$ when Pr is close to 100. 
\item
A change in behaviour can be clearly observed  in the 
 refined prediction  of the stability boundary that is visualised in the 
 rightmost plot. The results suggest that if Prandtl number  is reduced below 100  then 
 the transition to unsteady flow as a function of the Rayleigh number occurs
 significantly earlier. 
\end{itemize}

Confirmation of  these  observations \rbl{can be realised by} using an  {\it adaptive strategy}. 
The idea is  to refine the surrogate approximation by generating  additional data points in regions 
of parameter space to reduce the interpolation error where the bifurcation boundary is most uncertain.
\rbl{One such approach is developed in~\cite{singh25}, where
 the classifier output provides a probabilistic measure of
 flow stability, and the Shannon entropy of these predictions is employed as an uncertainty indicator. 
 A  flow-based deep generative machine learning strategy can then be trained to approximate a probability 
 density function that concentrates sampling in regions of high entropy, thereby directing computational effort
 toward the evolving bifurcation boundary. This coupling between classification and generative modelling 
 establishes a feedback-driven adaptive learning process analogous to error indicator based refinement 
 in contemporary PDE solution strategies.}

 \section{Concluding remarks and critical assessment} \label{sec:conclusions}

While reliably  trained neural networks offer the possibility of  transformative  technology in particular areas
of science, a convincing case for replacing physics-based modelling by contemporary machine-learning strategies  
is still to be established at this time.  The strategy presented herein has a very different flavour: the 
interpolating potential  of a  trained neural network is utilised in a role where it is
known to be state-of-the-art.  \rbl{The bifurcation tracking strategy is not perfect however. 
Here is a  list of factors that need to be considered before adopting it:
\begin{itemize}  \itemindent 0pt  \itemsep 2pt
\item 
Reproducibility is an obvious issue in machine learning. The loss function is
rarely  reduced to zero, even after intensive training. This means that  
``converged'' learned weights in the surrogate model will be different if
the training is repeated with a different realisation of the random initial weights. 
While this is a significant problem in a  deep learning setting, it is less of an issue 
when training  a shallow neural network with a relatively small number of parameters.
\item 
The physics needs to be faithfully represented by the discretised mathematical model if the 
results are to be  predictive in the sense of being able to reproduce 
results observed in practical experiments. Spatial approximation needs to be done carefully 
with either mesh refinement checks to confirm accuracy  or refinement determined  by 
reliable a posteriori error control. 
Our use of a  robust  adaptive time-stepping 
strategy to follow the evolution of a flow from a quiescent state to a final steady or  periodic state 
is a notable feature of our strategy. Integration of the Navier--Stokes equations in time has typically 
been done in the past using fixed time steps.  While fixed time-stepping methods can be  effective
 in modelling the growth or otherwise of perturbations of the equilibrium flow, they cannot be used to 
 model a  laboratory flow experiment from start to finish. 
 \item
 The identification of a sensible bifurcation detection strategy is problem dependent and 
 requires physical insight. For the three cases in this study,  prior testing of alternatives 
 was done to identify a  suitable measure of departure from the equilibrium state and
 inform the choice of tolerance for  labelling the training data.  The detection process
needs to  be robust  in the sense that the distinctive features of the  bifurcation picture 
are independent of the precise choice of tolerance.  This can be easily checked---one
simply needs to rerun the training process. The expensive step of generating the raw 
data does not need to be repeated.
 \end{itemize} }

While we have  limited our attention to two-dimensional  bifurcation problems in this work, our
 machine learning strategy generalises effectively to  high-dimensional parametric problems.
Moreover, while we have concentrated on tracking the {\it first} bifurcation point, there is
 no reason why the methodology cannot be to adapted to locate a succession of  branch points. 
All the  flow problems considered in this work  are representative of  problems where a classical linearised stability
analysis is predictive of reality in the sense of being able to reproduce results observed in  laboratory experiments.
 One interesting  future direction would be to consider applying the methodology  to flow problems where 
 transient growth of instabilities  limits the  applicability of classical eigenvalue analysis.  \rbl{Another
 interesting extension of this work would be to incorporate  labels generated from laboratory experiments
 into the training process.}

\section*{Reproducibility statement}
 The numerical results in this paper are all reproducible. The flow stability data was generated by running IFISS version 3.7 (\url{https://www.manchester.ac.uk/ifiss}) in batch mode. The IFISS software and user guide can be downloaded from the GitHub repository (\url{https://github.com/mcbssds/IFISS_download}). IFISS can be run using MATLAB (licensed by The MathWorks) or using the latest release of GNU Octave (available from \url{https://octave.org}). The data was processed using open-source software written in Python. The functions used to generate the figures in the paper and the MATLAB codes used to generate the batch files can all be downloaded from the GitHub paper repository: \url{https://github.com/mcbssds/hydrostabilityNN}.

\bibliographystyle{ACM-Reference-Format}
\bibliography{hydrostability}

\end{document}